\documentclass[]{aa}
\usepackage{graphicx}
\usepackage{subfigure}

\begin{document}
\title{The first detection of Far-Infrared emission associated with an
   extended HI disk\thanks{Based on
   observations with the {\it Infrared Space Observatory (ISO)}, an ESA project
   with instruments funded by ESA member states (especially the PI countries:
   France, Germany, the Netherlands, and the United Kingdom) and with
   the participation of ISAS and NASA.}. The case of NGC~891.}
\author{Cristina C. Popescu\inst{1,2}, Richard. J. Tuffs\inst{1}}

\offprints{Cristina.Popescu@mpi-hd.mpg.de}
\institute{Max-Planck-Institut f\"ur Kernphysik,
           Saupfercheckweg 1, D-69117 Heidelberg, Germany\\
\email{Cristina.Popescu@mpi-hd.mpg.de; Richard.Tuffs@mpi-hd.mpg.de}
           \and
           Research Associate, The Astronomical Institute of the Romanian
           Academy, Bucharest, Romania.}

\date{Received; accepted}

\abstract{Spiral galaxies in the local universe are commonly observed to be 
  embedded in extended disks of neutral hydrogen - the so called 
  ``extended HI disks''. Based on observations made using the ISOPHOT
  instrument on board the Infrared Space Observatory, we report 
  the first detection of cold dust in the extended HI disk of a spiral 
  galaxy. The detection was achieved through a dedicated deep Far-Infrared
  observation of a large field encompassing the entire HI disk of the edge-on 
  spiral galaxy NGC~891.
  Our discovery indicates that the extended HI disk of NGC~891 is not 
  primordial in origin.  

\keywords{galaxies: spiral, galaxies: structure, galaxies: evolution, 
ISM: dust, infrared: continuum}
}
\authorrunning{Popescu \& Tuffs 2003}
\titlerunning{The first detection of Far-infrared emission associated with an 
extended HI disk.}
\maketitle

\section{Introduction}

Radial gas surface density profiles in spiral galaxies show quite
similar behaviour in relation to the optical disk, irrespective of their
morphological type (Sancisi 1995, 1999). Whereas the molecular gas is 
concentrated towards the inner disk, the HI surface density is generally flat 
at an average level of $10\,{\rm M}_{\odot}\,{\rm pc}^{-2}$ over the 
entire extent of the optical disk (though with some variation from 
galaxy to galaxy, see Broeils \& van Woerden 1994). Exterior to the optical 
disk, which tends to have a comparatively abrupt cut-off at $\sim3$ stellar 
exponential scale lengths (Pohlen et al. 2000), the HI surface density falls 
off exponentially until a level of ca. $0.1\,{\rm M}_{\odot}\,{\rm pc}^{-2}$
is reached. At this point the gas disk either ends, or becomes ionised by the 
intergalactic medium. The portion of the HI disk extending beyond the optical 
stellar disk is commonly referred to as the ``extended HI disk''. 

  It is unknown 
  whether these gaseous disks are remnants of primordial material left over 
  from the epoch of galaxy formation (Larson 1990), or whether they 
  contain material reprocessed in stellar interiors, either transferred from 
  the
  stellar disk or captured from other galaxies. 
  In the latter case the extended HI disks should be enriched by metals 
  produced in stars (Tinsley \& Larson 1978, Pei, Fall \& Hauser 1999, Maller
  et al. 2001), so observations of these species could be 
  used to identify their nature.  Unfortunately metals in extended HI disks 
  are difficult to detect in the gas phase, either through their emission line
  spectrum (because of the lack of exciting stars) or through absorption 
  (because of the lack of sufficient background sources). However any metals
  present in form of dust grains offers an alternative way to trace the origin
  of these gaseous disks. Tentative evidence
for the presence of metals in the form of grains  was provided
from measurements of colour variations between background galaxies and control 
fields (Zaritsky 1994). There is also evidence that, within the confines of the
optical disk, grains have a larger scale length than the stars (Alton et
al. 1998a, Davies et al. 1999, Trewhella et al. 2000, Radovich, Kahanp\"a\"a \&
Lemke 2001, Xilouris et al. 1998, Xilouris et al. 1999, Popescu et al. 2000a,
Misiriotis et al. 2001) and that dust extends right up 
to the edge of the optical disks (Cuillandre et al. 2001). In some 
Blue Compact Dwarf galaxies observed in the FIR by Tuffs et al. (2002a,b) it 
has also been suggested that there is dust 
outside the optical emitting core region (Popescu et al. 2002).  

Here we present the first detection of cold dust in an extended HI disk, 
achieved 
through a dedicated deep Far-Infrared (FIR) observation of a large field 
encompassing the entire HI disk of the edge-on spiral galaxy NGC~891, 
made using the ISOPHOT instrument (Lemke et al. 1996) on board the Infrared
Space Observatory (ISO)(Kessler et al. 1996). We chose NGC~891 for this
observation as it has an asymmetric HI disk (Swaters et al. 1997), so 
any FIR counterpart should also be asymmetric and thus be more easily 
recognisable.

\section{Observations and data reduction}

A detailed description of the observations and data reduction is given in
Popescu et al. (2003). Here we give only a brief overview of these procedures.
The observations were made using the ISOPHOT-C200 2$\times2$ pixel array
in the C160 and C200 filters, which respectively cover passbands of
$130\,-\,218$ and $170\,-\,239\,\rm \mu m$ and have central
wavelengths of 170 and 200\,${\mu}$m. These FIR wavelengths were chosen since
they provide maximum sensitivity to any cold dust present in the extended HI
disk. Since there are no local heating sources in this region and the only 
available photons to heat the grains would be those coming from the optical 
disk, the FIR emission from any embedded dust was anticipated to be faint and 
have a spectral peak near our chosen filters. In order to cover the entire HI 
disk of NGC~891 (extending up to $\sim10$ arcmin from the nucleus; Swaters et
al. 1997), as 
well as the surrounding background, a field of radius $\pm13.5$ 
arcmin ($\pm 40$\,kpc) was mapped along the major axis of the galaxy. 

The FIR maps were obtained using the ``P32'' mapping mode which provided near 
Nyquist sampling over the three overlapping fields: north, south and
central. Preliminary results obtained on the central field were presented by
Popescu \& Tuffs (2002a), Tuffs \& Popescu (2003) and by Dupac et al. (2003). 
The data were processed using the 
latest P32 reduction package (Tuffs \& Gabriel 2003), which corrects 
for the transient response of the detector pixels. This allowed high 
dynamic range maps to be constructed to levels of 1 percent of the peak 
disk brightness. A time dependent flat field correction was made for 
each map, by fitting a cubic function to the response of the
detector pixels to the background. Calibration was made using V8.1 of
the ISOPHOT Interactive Analysis Package PIA (Gabriel et al. 1997). 

Radial profiles were derived from the background subtracted FIR maps by 
integrating the emission parallel to the minor axis of the galaxy in bins of 
width 18 arcsec along the major axis. Independent data contribute to each map 
pixel (and to each point on the profile)(Tuffs \& Gabriel 2003)

\section{Results}

\begin{figure}[htb]
\includegraphics[scale=0.75]{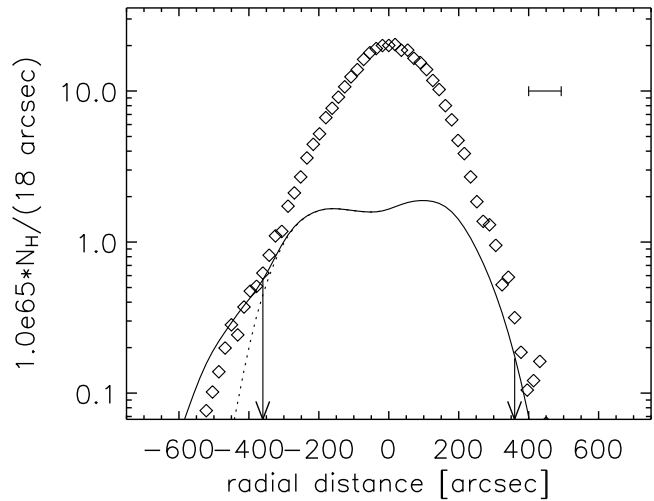}
\caption{The radial profiles of HI emission 
  convolved with the ISOPHOT PSF
  (solid line) and of 
  $200\,{\mu}$m FIR emission (symbols) sampled at intervals of
  18 arcsec. The negative radii correspond to the southern side of the
  galaxy and the galaxy was scanned at 60 degrees with respect to the major
  axis. The units of the FIR profile are W/Hz/pixel, multiplied with a
  factor of $2\times 10^{-22}$ and the error bars are smaller than the 
  symbols. The horizontal bar delineates the FWHM of the ISOPHOT 
  PSF of 93$^{\prime\prime}$. 
  The vertical arrows indicate the maximum extent of the optically
  emitting disk. The dotted line represents a modified HI profile 
  obtained in the southern side from the original one by cutting
  off its emission at the edge of the optical disk and by convolving
  it with the ISOPHOT PSF
    }
\end{figure}

NGC~891 was found to be a normal galaxy in respect to its integrated 
FIR properties, such that the fraction of stellar light reradiated by dust in 
the FIR is $\sim30\%$ (Popescu \& Tuffs 2002a), which is close to the mean
value for normal galaxies (Popescu \& Tuffs 2002b).
A detailed analysis of the FIR surface brightness distribution and 
profiles within the optical disk 
($\pm 360^{\prime\prime}$) has been presented in Popescu et al. (2003). There 
we find that the measured 
profiles and surface brightness distributions are in excellent agreement with 
the prediction for their counterparts obtained using the model for the 
optical/FIR/submm spectral energy distribution of Popescu et al. (2000a). The 
derived intrinsic distributions of dust and stars in NGC~891 were 
constrained from the optical/NIR images of NGC~891, as well as using
data at 60, 100, 450 \& 850\,${\mu}$m (Alton et al. 1998b) and data at 
1.3~mm (Guelin et al. 1993). Here we
compare the FIR profiles with the HI profiles and concentrate on the
emission beyond the optical disk.

In Fig.~1 we show the resulting radial profile at 200\,${\mu}$m, overlaid 
with the corresponding HI profile. The latter was obtained from the 
HI maps of Swaters et al. (1997) after convolution to the PSF of the
ISO measurements. Within $\pm200^{\prime\prime}$ from the 
centre,
where the HI radial profile is fairly flat, the 200\,${\mu}$m profile 
rises continuously towards the nucleus. This can be attributed in part to an 
increasing surface density of grains associated with molecular gas, which is
known to predominate in the inner disk (Garc\'{\i}a-Burillo \& Gu\'elin 1995),
as well as to the stronger radiation fields from the inner parts of the galaxy.
Between $\pm 200^{\prime\prime}$ and $\pm 360^{\prime\prime}$ (the edge of the
optical disk) both the FIR and the HI profiles fall
steeply. 
The $200\,{\mu}$m profile can be traced out to a radius of
$522^{\prime\prime}$ (24 kpc) in the South - $160^{\prime\prime}$ 
(7.4\,kpc) beyond the edge of the optical disk, and out to 
$432^{\prime\prime}$ (19.9 kpc) in the North - $70^{\prime\prime}$ beyond the
optical disk. By comparison the HI profile extends out to 
$600^{\prime\prime}$ in the South and out to $440^{\prime\prime}$ in the 
North (Swaters et al. 1997). Thus, the extent and the asymmetry in the
200\,${\mu}$m emission follows that of the HI emission (the same is true for the 
170\,${\mu}$m radial profile), indicating a dust emission 
counterpart to the extended HI disk. 

\begin{figure}[htb]
\hspace*{-2.0cm}
\includegraphics[scale=0.75]{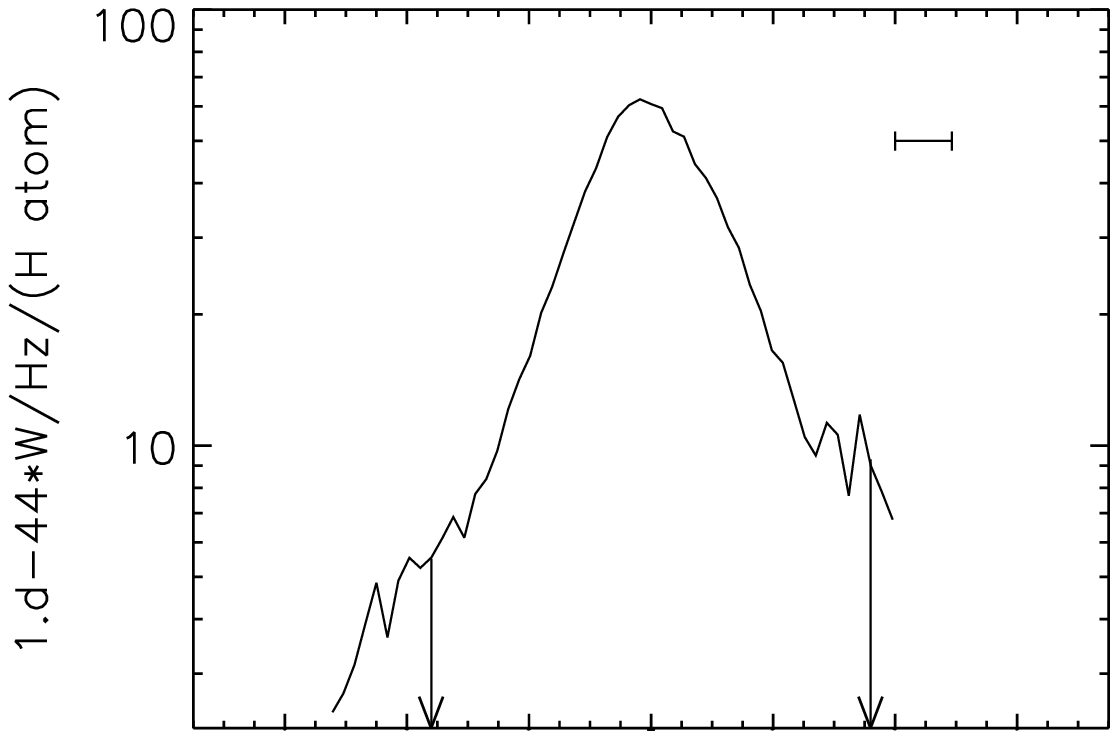}
\hspace*{0.4cm}
\includegraphics[scale=0.75]{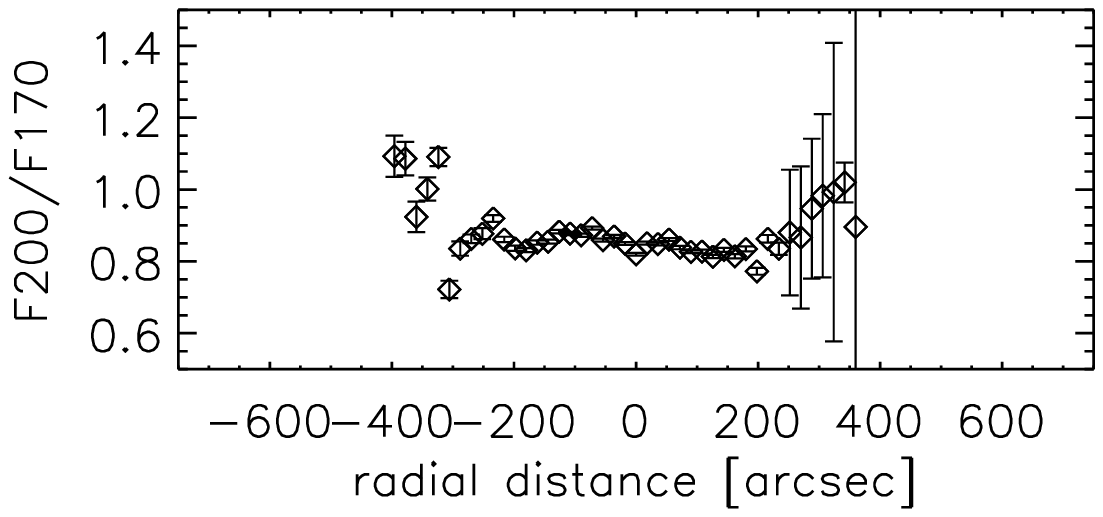}
\caption{Top panel: The radial profile of the ratio of the 
200\,${\mu}$m and HI emission. Again the vertical arrows indicate the 
maximum extent of the optically emitting disk and the horizontal bar 
delineates the FWHM of the ISO PSF of 93$^{\prime\prime}$.
Bottom panel: The radial colour profile F200/F170.
}
\end{figure}

In order to check that the FIR emission detected in the extended HI disk
is not attributable to beam smearing of emission from within the 
outer confines of the optical disk, we truncated the HI distribution at the 
edge of the optical disk (such that no emission would exist beyond 
this edge), convolved it with the ISOPHOT beam, and overplotted the 
result in Fig. 1. It is clear that the FIR emission detected (in the South) beyond the optical 
disk falls above this modified HI profile, indicating 
that the detection is real and not of an instrumental nature. 
We also note that the bright emission from the central region of the galaxy 
would not contribute to the emission seen towards the extended HI disk, 
as the ISOPHOT beam is known to fall steeply, and has no extended wings.

The relative rate at which the 200\,${\mu}$m and HI profiles
decline with radius is best seen in Fig.~2 (top panel), 
where the profile of the ratio of 
these quantities is plotted. On the southern side this ratio decreases 
by a factor of 2.3 between $360^{\prime\prime}$ (the edge of the optical disk) 
and  $522^{\prime\prime}$ (the maximum extent of the detected FIR emission). 
This could either be due to a
decrease in the dust-to-gas ratio, or to a decrease in the grain heating
with radius, or to a combination of the two. The dust-to-gas ratio can be
calculated from the ratio plotted in Fig.~2 (top panel), 
if the dust temperature 
$T_{\rm D}$ is known. In Fig.~2 (bottom panel) it is shown that the 
200/170 radial colour profile (measured as far as a radius of
$400^{\prime\prime}$) has a smooth progression towards colder emission
with increasing radial distance, which can be taken as evidence that the dust
at large galactic radii should be cold. From the 200/170 colour ratio 
of 1.1, as measured at a radius of $400^{\prime\prime}$, and assuming a grain 
emissivity proportional to ${\lambda}^{-2}$, we derive $T_{\rm D}=14$\,K.  
This provides us with an upper limit for $T_{\rm D}$ at radii larger than
$400^{\prime\prime}$, because the radiation fields should decrease continuously
with increasing distance from the edge of the optical disk. For the distance of
NGC~891 of 9.5\,Mpc (van der Kruit \& Searle 1981) and assuming a mixture of 
silicate and graphite 
grains (Laor \& Draine 1993, Draine \& Lee 1984), we derive a dust-to-gas 
mass ratio of 0.0083 at the 
projected radius of $400^{\prime\prime}$. This value is close to the 
dust-to-gas ratio of $\sim 1\%$ of both the local Galactic interstellar 
medium and of NGC~891 as a whole. Thus our measurements are consistent with 
there being little or no variation in the dust-to-gas ratio between the 
optical disk and the extended HI disk. So the radial fall-off in the ratio 
between the 200\,${\mu}$m and HI profiles in the extended HI disk is most 
probably due to the diminishing heating rate. This inference falls broadly in 
line with the expected decline in photon flux by a factor 2.1, if the
photons heating the grains originate from near the edge of the optical disk.

\section{Discussion and Conclusions}

The existence of large amounts of grains in the extended HI disk of
NGC~891 raises the challenging question about their origin and the 
implications for the origin of the extended HI disk itself. The large 
value of the dust-to-gas ratio
obtained for the extended HI disk clearly indicates that this gaseous disk is 
not primordial, left over from the epoch of galaxy formation. The detected
grains must have either been transported from the optical disk, 
or they must have been produced outside the galaxy. 

If the grains were transported from the optical disk continuously
over the lifetime of the galaxy, only a very small fraction of grains 
produced in the optical disk need to be transferred to explain the derived 
dust mass in the extended HI disk, since there are no obvious grain 
destructions mechanisms operating there. Taking the lifetime of NGC~891 to be 
$\tau_{\rm gal}=10^{10}$\,yr and 
the survival timescale of grains in the optical disk  
$\tau_{\rm surv}=10^8$\,yr, the total grain
production mass is $M_D^{\rm tot}= (\tau_{\rm gal}/\tau_{\rm surv})\times 
M_{\rm D}^{\rm opt}$, where $M_{\rm D}^{\rm opt}$ is the observed 
mass of grains in the optical disk at the current epoch (Popescu et al 2000a). 
We obtain 
$M_D^{\rm tot}= 1.3\times10^{10}\,M_{\odot}$. If we compare this mass 
with the total mass of dust present in the extended HI disk 
$M_{D}^{\rm ext}=2.3\times 10^6\,{\rm M}_{\odot}$ we can derive an 
efficiency of transfer of dust grains from the optical disk $\eta=1.8\times
10^{-4}$. Thus, it would only require a tiny amount of grains to be 
transported to the extended HI disk to account for the observations. 
This raises the possibility that the grains were transported via the prominent
halo of NGC~891, as originally traced in H$\alpha$ by Dettmar (1990) and
Rand et al. (1990). For the specific case of 
transporting grains into the halo several mechanisms have been proposed by 
Ferrara (1991), Davies et al. (1998) and Popescu et al. (2000b), though
no theory exists for the transport of grains through the halo to higher 
galactocentric radii.
However it would be a remarkable coincidence that the transfer efficiency
inferred from our observations should take 
exactly the value for which the present dust-to-gas ratio in the extended 
HI disk matches the dust-to-gas ratio in the optical
disk. Furthermore any transport of grains via the halo should produce a 
symmetrical distribution of dust, contrary with what is observed. 

An alternative mechanism for transporting grains and gas from the inner disk 
would be diffusion triggered by macro turbulence. To explain our observation, 
this mechanism would also have to be effective beyond the optical disk, in 
regions unperturbed by mechanical energy input from supernovae and stellar 
winds. A further requirement for this mechanism to be effective would be 
that the timescale for the mixing mechanism should be shorter than the 
timescale for grain destruction in the optical disk. As in the case of the
transport of grains via the halo, an argument against this mechanism is however
the observed asymmetry of the FIR profile. 

Another possibility is that both the gas and the dust in the extended HI disk 
were part of the interstellar medium of another galaxy which was (long ago) 
tidally stripped and captured by NGC~891\footnote{We note that NGC~891 is
thought to be a non-interacting system at the current epoch.}. This would 
also explain the 
asymmetry in both the HI and in the dust. A present day example
of such an interaction-accretion event
is the advanced interaction of a dwarf galaxy with M~101 (van der Hulst \&
Sancisi 1988). 

To conclude, while the exact mechanism through which the extended HI disk is 
formed remains unclear, our detection of FIR emission rules out a primordial 
origin of the extended HI disk in NGC~891. For the moment our
result was obtained for one galaxy and cannot therefore be generalised to all
spiral galaxies. However, future observations will be able to prove if 
in general extended HI disks of spiral galaxies contain large amounts of dust 
or if this is a characteristic peculiar to NGC~891. If the former is 
the case,
then this implies that at the epoch when the first galaxies formed, there 
must have been a rather efficient process which removed primordial debris 
from around the forming galaxies, for example in a strong galactic wind, or 
simply as a result of a rather efficient conversion of gas into stars.
Another implication of our detection of 
an asymmetric dust counterpart to the extended HI disk in NGC~891 is
that the asymmetry of the latter is intrinsic rather than being due to the
disk becoming ionised at a shorter radius.

\begin{acknowledgements}
We would like to thank the anonymous referee for his useful comments and
suggestions. We would also like to thank Andreas Burkert and Jay Gallagher 
for helpful discussions and Rob Swaters for providing us with the HI map.
\end{acknowledgements}

\end{document}